\documentclass[%
 prl,
 amsmath,amssymb,
 reprint,%
]{revtex4-1}

\usepackage{graphicx}
\usepackage{dcolumn}
\usepackage{bm}

\begin{document}

\title{Quantum Measurements Cannot be Proved to be Random}%

\author{Caroline Rogers}

\begin{abstract}
We show that it is impossible to prove that the outcome of a quantum measurement is random.
\end{abstract}
\maketitle

\section{Introduction}

The postulates of quantum mechanics say that when a measurement is made on a quantum state 
the outcome of the measurement is random \cite{nielsen}.
In classical physics, if the state of a closed system is known, then the state of that system is completely predictable.
As far as we know, randomness only exists in quantum mechanics \cite{gruska}. Everett proposed a purely unitary quantum theory \cite{everett} (without probability). However no known version of Everett's purely unitary quantum theory can account for the appearance of probabilities without extra ad hoc postulates \cite{ken09}. The probabilities in quantum mechanics are thought to be caused by randomness \cite{nielsen}. In this letter we show that, given a sufficiently large sequence of experimental data, there is no proof that the data is random \cite{vitanyi}. Thus it is actually impossible to prove that randomness exists in quantum mechanics.
\\

What is randomness? The outcomes of quantum measurements appear to be random so first we shall discuss how numbers can appear to be random when they are not random. 

Consider SHA-1, the Secure Hash Algorithm \cite{nsa02}. SHA-1 was designed by the National Security Agency of the United States for cryptographic purposes but can be used as a random number generator. A random number generator does not actually generate random numbers \cite{dia02}. 
\begin{quotation}
Anyone who considers arithmetical methods for producing random digits is, of course, in a state of sin. [von Neumann] \cite{vitanyi}
\end{quotation}
A random number generator is a deterministic algorithm which, given a particular seed, always produces the same sequence of ``random" numbers. The sequence of numbers generated by a random number generator is often called ``random" because the sequence appears to be random.

SHA-1 was designed to prevent even a supercomputer from finding the seed that generated a sequence of ``random" numbers. SHA-1 was designed so that no one could find out that the sequence of ``random" numbers generated is not random without knowing the seed. The ``random" numbers generated by SHA-1 appear to be random even though they are not random \footnote{In 2005, a flaw was discovered in the SHA-1 hash algorithm \cite{sch05}. SHA-1 was previously thought to generate a sequence of numbers which no one could discover were not random. Now SHA-2 \cite{nsa02} is recommended instead.}.

Commercial organisations make quantum random number generators \cite{idquantique} which generate random numbers by measuring quantum states. It is widely believed that the numbers generated by a quantum random number generator are truly random \cite{nielsen}. We are going to show that there can be no proof that a quantum random number generator generates truly random numbers. To do this, we still need to figure out what randomness is.
\\

The notion of randomness is not strict \cite{vitanyi}. We cannot say that 1,000,000 is not a random number but 1,000,001 is a random number. In statistics one cannot say that a hypothesis holds for certain so one says that one is e.g. 95\% certain that a hypothesis holds. We can do the same for randomness and say that a sequence $x_1x_2\ldots x_n$ is $c$-random (rather than just random). To find out what precisely is $c$-random, we can look at what is not random.

Suppose that a quantum random number generator generates a random sequence $x=x_1x_2\ldots x_n$ of $n$ bits. Suppose that $x$ is the first $n$ digits of the binary expansion of $\pi$. If $n$ is large we can be sure that $x$ is not random and that the quantum random number generator is not generating random numbers. 

Suppose again that a quantum random number generator generates a random sequence $x=x_1x_2\ldots x_n$ of $n$ bits. Now suppose that there is a hidden deterministic algorithm $A$ that gives the value of each bit $x_i$ of $x$. To keep things simple, suppose that all the measurements on the quantum state that generates $x_1x_2\ldots x_n$ are made simultaneously. Assuming that all the measurements are made simultaneously makes $A$ more powerful because $A$ can see that the output is of length $n$ even when the first bit $x_1$ is being created. We make $A$ more powerful because we want to be sure when we say that a string $x$ is very random that $x$ has not been generated by a deterministic algorithm. $A$ is a deterministic but hidden algorithm that creates $x$ on input $n$.

Given a sufficient but finite amount of time, a computer program $T$ can find $A$ by looking for every possible algorithm $B$ that could have created $x$ on input $n$. Segre \cite{seg09} has looked to see if there are such algorithms on a real computer. Some algorithms $B$ might go round in circles and not halt, so to find $A$, $T$ does the following steps. 

In the first step $T$ runs all the algorithms $B$ of length $1$ for $1$ step on input $n$. In the second step, $T$ runs all the algorithms $B$ of length $2$ for $1$ step on input $n$ and all the algorithms $B$ of length $1$ for $2$ steps on input $n$. In the $i$th step, $T$ runs all the algorithms of length $i$ for $1$ step on input $n$, all the algorithms of length $i-1$ for $2$ steps on input $n$, $\ldots$ and all the algorithms of length $1$ for $i$ steps on input $n$. Whenever $T$ finds a new algorithm $B$ which outputs $x$ on input $n$, $T$ prints out the algorithm $B$. 
In this way, $T$ will eventually print out $A$.

If $A$ has a very short description such as ``$x=x_1x_2\ldots x_n$ and $x_i$ is the $i$th binary digit of $\pi$", then $x$ is not at all random. If $x$ is truly random in the sense that $x$ has been generated by a true random variable $X$ with $P(X=0)=P(X=1)=1/2$, then we would be very surprised if there is a simple deterministic algorithm $A$ which given $n$ describes $x$. This gives us our measure of $c$-randomness.

We say that $x=x_1x_2\ldots x_n$ is $0$-random when there is no deterministic algorithm $A$ that describes $x$ apart from $x$ itself. If $x$ is $0$-random and $n$ is large, then we can be sure that $x$ is actually random and there is no hidden deterministic algorithm $A$ that creates $x$.
\\

We now define $c$-randomness. Let us define $C(x|l(x))$ to be the length of the shortest algorithm $A$ which outputs the  string $x=x_1x_2\ldots x_n$ given the length $l(x)=n$ of $x$ in bits as input. $C(x|l(x))$ is the length of the shortest description $A$ of a string $x=x_1x_2\ldots x_n$ given the length $l(x)$ of $x$ as input. I.e.
\[
C(x|l(x)) = \min_{U(l(x),A)=x}l(A) 
\]
where $U$ is a fixed universal programming language such as C or FORTRAN and $U(l(x),A)=x$ means that the universal programming language outputs $x$ on input $l(x)$ and $A$. The quantity $C(x|l(x))$ is known as the Kolmogorov complexity of $x$ given $l(x)$ \cite{chaitin,kolmogorov,solomonoff1,solomonoff2}.

The randomness of a string $x$ increases as the length of the shortest deterministic algorithm $A$ which outputs $x$ on input $n$ increases. $x$ is $0$-random when the length of the shortest deterministic algorithm $A$ which outputs $x$ on input $n$ has at least the same length as $x$. We can formally define $c$-randomness by defining $\delta_0$, the Martin--L\"{o}f reference universal test for randomness with respect to the uniform distribution $L$ \cite{vitanyi,martinlof}:
\[
\delta_0(x|L) = l(x)-C(x|l(x))-1 \; .
\]
The randomness of $x$ increases as $\delta_0(x|L)$ decreases. When $\delta_0(x|l(x))=n/2$, there might be an algorithm $A$ of length $n/2$ that creates $x$ or there might be a mix of randomness and determinism for creating $x$. We are not sure. 

We say that $x$ is $c$-random when $\delta_0(x|L)\leq c$. If $x$ is $0$-random and $n$ is large, we are very confident that there is no deterministic algorithm $A$ that creates $x$ on input $n$ (unless the length of $A$ is at least as long as $x$ itself). If there is no deterministic algorithm $A$ shorter than $x$ which creates $x$ on input $n$, then $x$ is random. If $n$ is large and $x$ is random, then we can be confident that a quantum random number generator that generates $x$ is truly random.
\\

We now discuss the main result of this letter: that it is impossible to prove that the outcome of a quantum measurement is random. 

Consider proving that the outcome of a quantum measurement is random. To prove that the outcome of a quantum measurement is random, we need to show that when $n$ is large there is a proof that a string $x$ generated by a quantum random number generator is $0$-random. To prove that the quantum random number generator is truly random, one should take $n$ as large as possible and prove that $x=x_1x_2\ldots x_n$ is $0$-random.  To prove that $x$ is $0$-random, we need to prove that $C(x|l(x))\geq n-1$. We will show that $x$ cannot be proved to be $0$-random when $n$ is very large because there is an integer $N$ such that there is no proof that $C(y|l(y))>N$ for any string $y$ \cite{vitanyi}.

We now prove the main result. We prove that there is no proof that the outcomes quantum measurements are random i.e. we prove that there exists $N$ such that there is no proof that $C(x|l(x))>N$ for any $x$. 

Let us fix $N$. Let $x^N$ be a string such that there exists a proof $P^N$ such that $C(x^N|l(x^N))>N$ (please note that $x^N$ and $P^N$ might not exist). The Church--Turing thesis \cite{church,turing} says that any reasonable calculation can be made by a computer. If the proof $P^N$ exists then there exists a computer program which can implement $P^N$ and prove that $C(x^N|l(x^N))>N$. 

On input $N$, a computer program $T'$ can search for a string $x^N$ and a proof $P^N$ such that $C(x^N|l(x^N))>N$ in the same way that the computer program $T$ searched for $A$ above. Some of the proofs $P^N$ that there is a string $x^N$ such that $C(x^N|l(x^N))>N$ might go round in circles and not halt so each proof $P^N$ is tried for a finite but increasing number of steps. A proof halts in a finite number of steps so some proof $P^N$ will be found if it exists.

In the first step, $T'$ tries all proofs $P^N$ of length $1$ for $1$ step and all strings $x^N$ of length $1$ to see if $P^N$ proves that $C(x^N|l(x^N))>N$. In the $i$th step, $T'$ tries all proofs $P^N$ of length $1$ and all strings $x^N$ of length $1$ to see if $P^N$ proves $C(x^N|l(x^N))>N$ in $i$ steps, $T'$ tries all proofs $P^N$ of length $2$ and all strings $x^N$ of length $2$ to see if $P^N$ proves $C(x^N|l(x^N))>N$ in $i-1$ steps, $\ldots$, $T'$ tries all proofs $P^N$ of length $i$ and all strings $x^N$ of length $i$ to see if $P^N$ proves $C(x^N|l(x^N))>N$ in $1$ step. If $T'$ finds a proof $P^N$ and a string $x^N$ such that $P^N$ proves that $C(x^N|l(x^N))$ then $T'$ outputs $x^N$ and halts. 

$C(x^N|l(x^N))$ is defined to be the length of the shortest description of $x^N$  given $l(x^N)$ as input in the fixed universal programming language. $T'$ outputs $x^N$ on input $N$ so $T'$ and $N$ together form a description of $x^N$ in the universal programming language. $T'$ is independent of $N$ and $x^N$ so the length of describing $T'$ and $N$ is $\log(N)+c$ bits where $c$ is a constant independent of $N$ and $y^N$. By the definition of $C(x^N|l(x^N))$ we have
\[
C(x^N|l(x^N)) \leq \log(N)+c  \; .
\]
However, by definition of $T'$, we have
\[
C(x^N|l(x^N)) > N \; .
\]
Putting these two inequalities together we have
\[
\log(N)+c < N 
\]
where $c$ is a constant independent of $N$. There is a limit on the size of $N$ that satisfies this last inequality. Thus we have proved our main result.

\section{Conclusions}

We have shown that it is impossible to prove that randomness exists in quantum mechanics. This is a fundamental result. 
Here are some ideas for future research.

Quantum cryptography is a commercial application of quantum mechanics \cite{nielsen}. It might be impossible to prove that some of the quantum cryptography protocols are secure if they rely on the outcomes of measurements being random.

The main result of this letter is an application of Kolmogorov complexity. Kolmogorov complexity has many applications  from pure mathematics \cite{cal99} to physics \cite{cav90,zur99}. Li and Vitanyi \cite{vitanyi} review the current applications of Kolmogorov complexity. Kolmogorov complexity has already been applied to quantum computation and generalized to quantum mechanics \cite{ber01,gacs,lap04,lap05,mor05,mor06,mor07,markusthesis,mue06,mue08,mue09,rogers,rog08a,rog08b,sch01,segre,seg09,svo96,tad02,tad06,vit01}.
An avenue for future research is to find more applications of Kolmogorov complexity to quantum computation and to find more applications of the quantum generalizations of Kolmogorov complexity.

In this letter, we have assumed that any hidden algorithm for creating the results of quantum measurements can be performed by a computer. Some might believe that this assumption might not be correct.
\begin{quotation}
Collectives generated by nature, as postulated by von Mises, may very well satisfy stricter criteria of randomness. Why should collectives generated by quantum mechanical phenomena care about mathematical notions of computability? Again, satisfaction of all effectively testable prerequisites for randomness is some form of regularity. Maybe nature is more lawless than adhering strictly to regularities imposed by the statistics of randomness. [Li and Vitanyi] \cite{vitanyi}
\end{quotation}

The Copenhagen interpretation explicitly states that when a quantum state is measured, the state collapses randomly onto one of the possible outcomes of the measurement. We have shown that it is impossible to confirm that this collapse is random, thus the main result of this letter might favour other interpretations of quantum mechanics. There is still scope for new classes of interpretations of quantum mechanics \cite{ken07} which this letter might motivate.

\section{Acknowledgements}

I thank Sougato Bose for drawing my attention to the work of Adrian Kent which motivated the writing of this letter. I thank Michael Nielsen for showing me how examples can be used to explain concepts easily.

I thank Mark Hinton for proof-reading this letter, for his comments and for his encouragement. I thank Mark Hinton for pointing out that probability does not always imply randomness because probability can be used to describe unknown but deterministic situations.

I thank Markus M\"{u}ller, Iain Stewart and Vlatko Vedral with each of whom I have had many interesting discussions about Kolmogorov complexity and quantum mechanics. I thank Markus M\"{u}ller for proof-reading my PhD thesis which contained a statement of the main results. I thank Markus M\"{u}ller for answering my question about where I can obtain experimental data from quantum measurements which is how I know of quantum random number generators.

I thank the Engineering and Physical Sciences Research Council for funding.

\bibliographystyle{prsty}
\bibliography{QuantumMeasurementsCannotBeProvedToBeRandom}

\end{document}